\documentclass[%
 aip,
 jmp,%
 amsmath,amssymb,
 reprint,%
]{revtex4-1}
\usepackage{graphicx}
\usepackage{dcolumn}
\usepackage{bm}
\usepackage[mathlines]{lineno}

\usepackage{graphicx}
\usepackage{times}
\usepackage{verbatim}
\usepackage{color}
\usepackage{longtable}

\begin{document}

\preprint{AIP/123-QED}

\title[]{Calibration of the Air Shower Energy Scale of the Water and Air Cherenkov Techniques in the LHAASO experiment}

\author{
F. Aharonian$^{26,27}$,
Q. An$^{4,5}$,
Axikegu$^{20}$,
L.X. Bai$^{21}$,
Y.X. Bai$^{1,3}$,
Y.W. Bao$^{15}$,
D. Bastieri$^{10}$,
X.J. Bi$^{1,2,3}$,
\\
Y.J. Bi$^{1,3}$,
H. Cai$^{23}$
% ...
\\
(The LHAASO Collaboration)
}

\affiliation{
	$^1$ Key Laboratory of Particle Astrophyics \& Experimental Physics Division \& Computing Center, Institute of High Energy Physics, Chinese Academy of Sciences, 100049 Beijing, China\\
	$^2$University of Chinese Academy of Sciences, 100049 Beijing, China\\
	$^3$TIANFU Cosmic Ray Research Center, Chengdu, Sichuan,  China\\
	$^4$State Key Laboratory of Particle Detection and Electronics, China\\
	$^5$University of Science and Technology of China, 230026 Hefei, Anhui, China\\
% ...
$^{10}$Center for Astrophysics, Guangzhou University, 510006 Guangzhou, Guangdong, China\\
% ...
$^{15}$School of Astronomy and Space Science, Nanjing University, 210023 Nanjing, Jiangsu, China\\
% ...
$^{20}$School of Physical Science and Technology \&  School of Information Science and Technology, Southwest Jiaotong University, 610031 Chengdu, Sichuan, China\\
	$^{21}$College of Physics, Sichuan University, 610065 Chengdu, Sichuan, China\\
% ...
$^{23}$School of Physics and Technology, Wuhan University, 430072 Wuhan, Hubei, China\\
% ...
$^{26}$College of Sciences, Northeastern University, 110819 Shenyang, Liaoning, China\\
	$^{27}$Dublin Institute for Advanced Studies, 31 Fitzwilliam Place, 2 Dublin, Ireland \\
% ...
}
\date{\today}
\email{caozh@ihep.ac.cn;zengzk@ihep.ac.cn;llma@ihep.ac.cn;\\
wangyanjin@ihep.ac.cn;nanyc@ihep.ac.cn}
\begin{abstract}
The Wide Field-of-View Cherenkov Telescope Array (WFCTA) and the Water Cherenkov Detector Arrays (WCDA) of LHAASO are designed to work in combination for measuring the energy spectra of various cosmic ray species over a very wide energy range from a few TeV to 10 PeV. 
The energy calibration of WCDA can be achieved with a proven technique of measuring the westward shift of the Moon shadow of galactic cosmic rays due to the  geomagnetic field. This deflection angle $\Delta$ is inversely proportional to the energy of the cosmic rays. 
The precise measurements of the shifts by WCDA allows us to calibrate its energy scale for  energies as high as 35 TeV. The energy scale measured by WCDA can be used to cross calibrate the energy reconstructed by WFCTA, which spans the whole energy range up to 10 PeV. In this work, we will demonstrate the feasibility of the method using the data collected from April 2019 to January 2020 by the WFCTA array and WCDA-1 detector, the first of the three water Cherenkov ponds, already commissioned at LHAASO site. 
\end{abstract}

\keywords{LHAASO, WCDA, WFCTA, Moon Shadow, Energy Scale, composition}
\maketitle
%
%
\section{Introduction}

Cosmic ray experiments based on the extensive air shower (EAS) technique usually feature different types of ground based detectors, such as a scintillation counter array, water Cherenkov detectors, or imaging air Cherenkov telescopes, each measuring shower properties in different ways. 
Therefore, it is mandatory to establish a way to calibrate the shower energy measurement between the different detectors, a non-trivial task given that it has to be done directly using cosmic ray data. 
Many experiments, such as ARGO-YBJ~\cite{ARGO-YBJ}, have successfully their energy-scale calibration by using the deficit of cosmic rays blocked by the Moon, as it moves inside the field of view (FoV) of the detectors, usually referred to as the Moon shadow of cosmic rays. 
The geo-magnetic-field (GMF), deflecting the charged cosmic rays, shift the Moon shadow on the ground with respect to the Moon real position.
The displacement of the shadow is clearly dependent on the cosmic ray rigidity and becomes negligible at high energies.
However, at energies below 40 TeV the shadow shift westward is clearly observable. 

\begin{figure}[ht]
\centering
\includegraphics[scale=0.30]{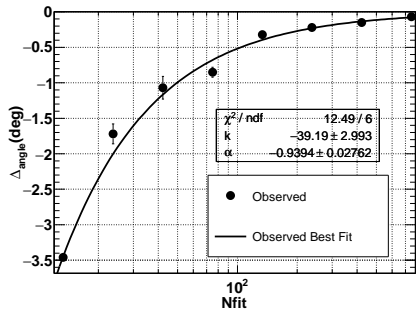}
\caption{The displacement of the Moon shadow from the real direction of the Moon as a function $N_{fit}$. }
\label{moon-shadows}       
\end{figure}

The Water Cherenkov Detector Array (WCDA) of the Large High Altitude Air Shower Observatory (LHAASO) has a detection threshold of about 1 TeV for cosmic rays.
WCDA has measured the Moon shadow shifts as a function of $N_{fit}$. Here, $N_{fit}$ is the number of triggered units with trigger time within 30 ns from the conical front. Units are the 5 m $\times$ 5 m cells in which the pond is subdivided.
A data set are collected from 2019/05/01 to 2020/01/31 with 731.2 hours of the Moon observation with zenith angle smaller than $45^{\circ}$. The reconstruction of arriving directions and shower cores can be seen in the reference ~\cite{WCDA-on-Crab}. During the calculation of the significance, events with cores located both inside and outside the pond are used. The significance of the Moon shadow for each $N_{fit}$ group is greater than $10\sigma$, enabling a good measurement of both the location and the width of the distribution. The deflection angle of the shadow as a function of $N_{fit}$ is plotted in Fig.~\ref{moon-shadows}. The  measured deflection can be approximated by the simple form $\Delta = k\cdot N_{fit}^{\alpha}$, where $k =-39.2\pm2.9$ and  $\alpha =-0.939\pm0.028$.   

At the same time, the shadow shift can be evaluated using a ray tracing algorithm with a detailed GMF model~\cite{GMF-model} and for different cosmic ray compositions, i.e., a mixture with different ratios of protons and alpha particles as a function of the their energies. 
The deflection angle measured by the Moon shadow shift provides a means to correlate the energy estimator and the energy of primary particles, thus providing a simple method for measuring the energy of the showers. In the energy range below 10 TeV, $N_{fit}$ is selected as the shower energy estimate\cite{WCDA-on-Crab}. Each group of events in $N_{fit}$ bins is assigned for the energies corresponding to the shifts just like what have done in ARGO-YBJ experiment\cite{ARGO-YBJ}.

The Wide FoV Cherenkov Telescope Array (WFCTA) of LHAASO measures shower characteristics using the Cherenkov light generated by secondary particles produced in the air during the whole shower development. 
The faint Cherenkov light is overwhelmed by moonlight, and can not be detected near the direction of the Moon, making it impossible  to calibrate the energy scale by measuring the Moon shadow shift. Nevertheless, we can use events that trigger both WCDA and WFCTA, to bridge the energy scale of WCDA to the WFCTA measurements. 
Since the trigger threshold of WFCTA is higher than 
for WCDA, only events with energies above 15 TeV can be used for the cross-calibration.  
At such high energies, the energy estimator of $N_{fit}$ for WCDA-1 is found saturated because of the natural limit of the 900 units of the detector. A more adequate energy estimator than $N_{fit}$ has to be defined first. The real challenge lies in the fact that the shift of the Moon shadow becomes small at such high energies and, at the same time, the number of events for this measurement decreases as energy increases. 
Therefore,  at this early stage of the LHAASO experiment, the limited statistics is the dominant uncertainty in this calibration procedure. Clearly,  accumulating data over time,  
the statistical error will be reduced until it becomes less important than the systematic uncertainties due to the unknown composition of the cosmic rays and the hadronic interaction models used to simulate showers. 
The overall uncertainty of the energy scale should become lower than 10\% after four years of observation.

In this work, firstly we briefly describe the LHAASO Observatory detectors in section~{\ref{sec:lhaaso-detector}}. In section~{\ref{sec:stats}}, we discuss how the energy scale for WCDA can be established using the Moon shadow shift measurement for energies above 6 TeV. The uncertainties of the energy scale are also discussed in the section. 

How the WCDA energy scale is propagated to WFCTA for calibrating the reconstruction of shower energies is described in section~{\ref{sec:WFCTA}} and, finally, in section~{\ref{sec:conclusions}} we draw some conclusions.

\section{The LHAASO WCDA and WFCTA arrays}\label{sec:lhaaso-detector}
The LHAASO Observatory is based on the so-called 'hybrid' approach for the measurement of shower characteristics, consisting in the simultaneous detection of atmospheric showers with different types of detectors.
The observatory is built around the three ponds of water Cherenkov Detector Array (WCDA), featuring 3120 gapless detecting units to instrument an area of 78,000 m$^2$. Near WCDA 18 wide field of view Cherenkov telescopes (WFCTA) are installed. They survey the sky above the whole array with a coverage of 4608 square degrees.
This core of the array is surrounded by 5195 scintillation counters (ED) and 1188 muon detectors (MD), which constitute an array covering an area of 1 km$^2$ (KM2A). The construction of LHAASO has nearly been completed. 

\begin{figure}[ht]
\centering
\includegraphics[scale=0.18]{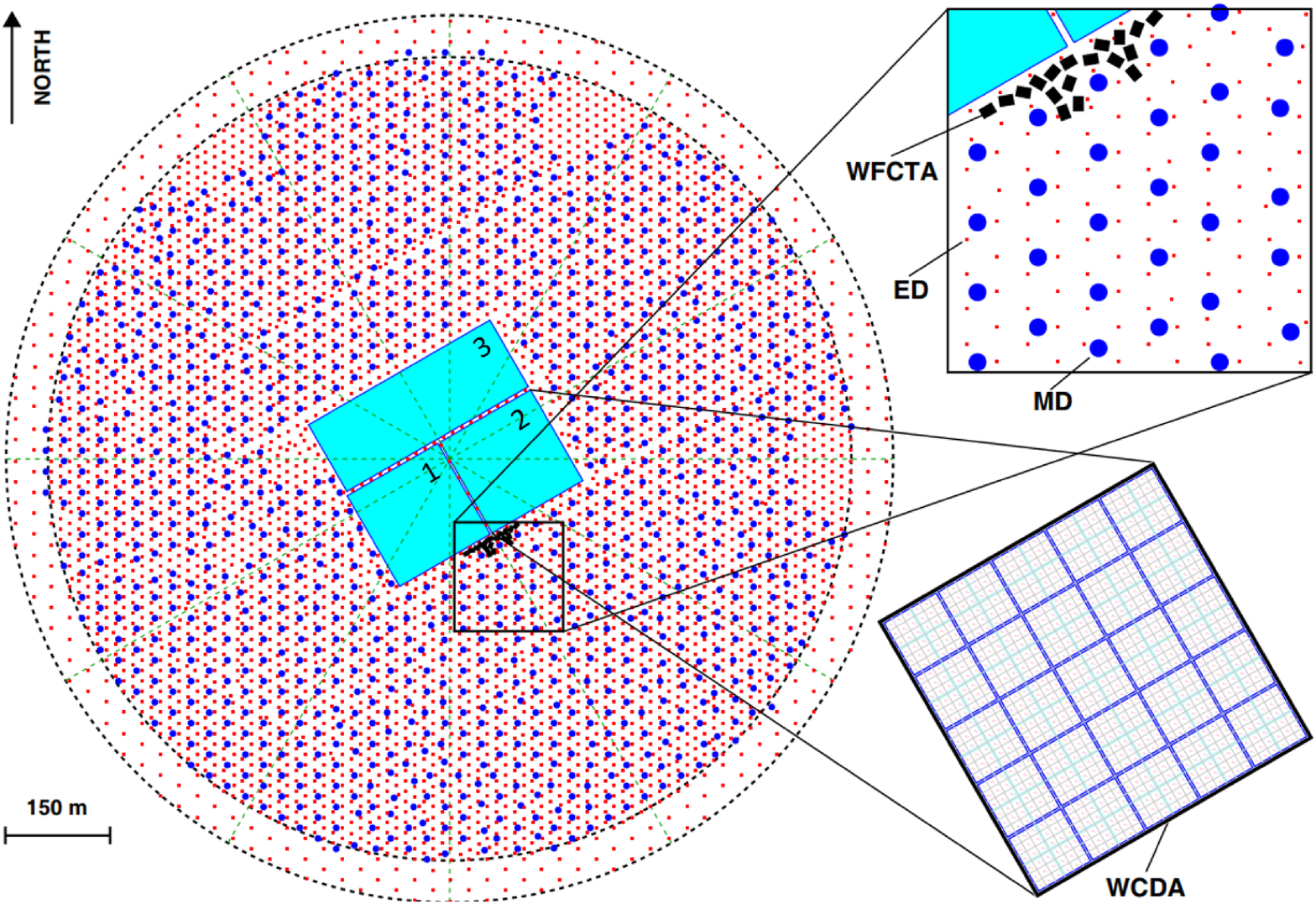}
\caption{The LHAASO layout. The three ponds of WCDA are represented by the cyan rectangles at the center of the site, each divided into square cells of 5 $\times$ 5 m$^{2}$ (as can be seen in the zoomed view at the bottom right). The 18 WFCTA telescopes, positioned near the WCDA, can be seen in the zoomed view at the top right. The remaining KM2A extends over an area of about 1 km$^2$, instrumented with  the electromagnetic detector (ED) array of  scintillation counters (small red dots) and the muon detector array (big blue dots).}
\label{layout}       
\end{figure}

 As shown in Fig.~\ref{layout}, WCDA is composed of two ponds with an area of 150 m $\times$ 150 m each and a third larger one with an area of 300 m $\times$ 110 m.
The smaller pond in the South-West direction, named WCDA-1, has started science operations since April 2019. It has 900 units, or cells, of 25 m$^2$, each equipped with a large (8") PMT used also for timing and a small (1.5") PMT at the center of each unit at 4 m of depth from the water surface. The use of two PMTs watching upwards allows us to cover a wide dynamic range spanning from 10 to 30,000 photo-electrons, which are produced in water by the shower charged secondary particles.
To suppress the light cross-talking effect and improve the timing resolution, black plastic curtains delimit the units. The front-end electronics (FEE) of the large PMTs is designed to achieve a time resolution of 0.5 ns, which together with the large dynamic range, enables the measurement of the particle density distribution in the shower cores without any saturation even for energetic showers up to 10 PeV. This allows to achieve determination of the  core location with a precision better than 3 m over a wide  energy range. 
WCDA-1 can measure shower directions with a resolution better than 0.2$^\circ$ above 10 TeV and 1.0$^\circ$ above 600 GeV~\cite{WCDA-on-Crab}.  The water transmission can be monitored by the "muon" peak observed by each unit~\cite{WCDA-on-Crab}. 

Six WFCTA Cherenkov telescopes in the southwest corner of WCDA-1, and two more in the southeast corner were put into science operation during  winter 2020. The centers of the six telescopes are located 100 m south and 38.5 m west of the wcda-1 center.
As shown in Fig.~\ref{layout}, the final layout will include 18 telescopes.
The telescopes have mirrors with an area of about 5 m$^2$,  composed of aluminized spherical mirror facets.
Each telescope can be tilted up and down in elevation from 0$^\circ$ to 90$^\circ$. 
A camera with 32 $\times$ 32 square pixels, realizing a FoV of 16$^\circ \times $16$^\circ$, is in the focal plane of the telescope, at 2870 mm away from the mirror center.
Each  pixel is realised by a silicon photomultiplier (SiPM) coupled to a square-surface light-funnel of 1.5 cm $\times$ 1.5 cm, corresponding to a FoV of 0.5$^\circ \times$ 0.5$^\circ$.  
Each SiPM is composed by an array of 0.36 million avalanche photo diodes (APDs) with a micro-cell size of 25 $\mu$m, covering a dynamic range from 10 to 40,000  photo-electrons with a measured non-linearity less than 5\%~\cite{SiPM}. 
In front of the photo-sensors and light-funnels, a window  is installed. The window is coated with a wide-band filter to suppress the light above 550 nm, where the night sky background light (NSB) dominates with respect to the Cherenkov signal. This is relevant since the photo-detection of SiPMs is considerably larger than PMT one at these wavelengths. The performances of SiPM are affected by the temperature and night sky background~\cite{SiPM-performance}. A calibrated led is installed in front of the camera to calibrate the performance of SiPMs in real time.   

\begin{figure}[h]
\centering
\includegraphics[scale=0.20]{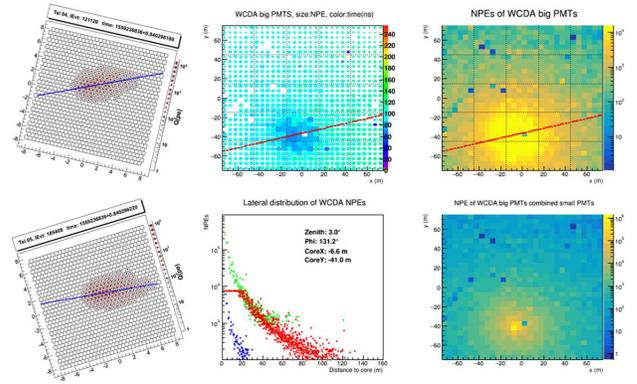}
\caption{A typical event that triggered both WCDA and WFCTA. 
The two tilted panels on the left-hand side show the images recorded by two of the WFCTA telescopes, both in vertical view, but with different orientations of $15^\circ$ and $45^\circ$, respectively, from North towards West. 
The two panels on the right, instead, show the maps of hits in WCDA units, where the color scale indicates the  number of recorded photo-electrons ($N_{pe}$).
The top right panel shows the map for the 8" PMTs, which exhibit a clear saturation in the central region of the shower.
The lower right panel shows the map for the same event as recorded by the 1.5" PMTs. 
The corresponding lateral distributions of $N_{pe}$ are shown in the lower panel in the middle column. Here, the red dots represent the 8" PMT and the blue ones the 1.5" PMTs,  which are also scaled (green dots) and overlaid on the previous one to reconstruct the profile over the full range. The upper panel shows the arrival time (in ns) of shower secondary particles as recorded by the WCDA-1.}
\label{event}       
\end{figure}

Each group of 16 pixels is clustered together in a FEE board connected with the readout system. Each pixel has a high-gain and a low-gain channel, each equipped with a 50 MHz 12-bit flash ADC to digitize the waveform. These two channels allow to cover the whole dynamic range of the SiPMs. 
A trigger signal, $T_0$, is generated by the high-gain channel of each pixel and transmitted to the trigger logic that collects the signals from all pixels.  
A pattern recognition algorithm is used to decide whether or not a shower has been observed. It generates a signal  $T_1$, which starts the readout of the digital waveform from every pixel. 
The total charge and the  average time of the waveform are transmitted to the LHAASO data center with an absolute time stamp. Further off-line triggers, in particular the inter-telescope trigger, and noise pixel rejection are carried out on the CPUs of the data center.
A typical common trigger event with WCDA-1 and two telescopes is shown in Fig.~\ref{event}.

\section{Determination of Shower Energy Scale with WCDA-1}\label{sec:stats}

The energy region where WCDA-1 and WFCTA can observe the same showers, at the maximum energy range for WCDA due to the higher energy threshold of WFCTA. In the energy range, as an energy estimator $N_{fit}$, the saturation appears, so a better energy estimator than $N_{fit}$ is needed for the energy scale determination. The energy scale of WCDA-1 can be obtained as following setps:

1: Selecting total number of detected photo-electron by PMTs, $N_{pe}$, as the energy estimator. The Moon shadow shift westward is measured as a function of $N_{pe}$, i.e., $\Delta$ vs. $N_{pe}$.

2: Calculating the Moon shadow shifts by tracing the cosmic rays with certain compositions through GMF as functions of CR promary energy, i.e. $\Delta$ vs. energy. 

3: Investigating the composition of cosmic ray events that trigger WCDA-1 in the relevant ranges of $N_{pe}$ using the air shower and detector response simulation. The composition is used as an input to establish the adequate relation between the measured $\Delta$ and the primary energy.  

4:  Investigating the effect due to the shower energy resolution and the power-law like CR spectrum to the match between $N_{pe}$ and the primary energy according to the corresponding Moon shadow shift. Establishing the energy scale, namely $N_{pe}$ versus $E_{median}$ in each group of $N_{pe}$. 


\subsection{$N_{pe}$ as the energy estimator for high energy showers}
WCDA measures Cherenkov photons generated by the secondary charged particles of showers, mainly $e^\pm$ and $\gamma$-rays, generated in air showers. The summed number of photo-electrons recorded by the PMTs is dependent not only on the number of secondaries, but also on their energies. Because the energy of the secondary particle in the air shower is proportional to the number of electrons in its induced shower in water, the total number of photo-electrons measured by units with trigger time within 30 ns from conical plane, denoted as $N_{pe}$, is a primary energy estimator. The simple form relating the primary energy and the estimator $E = b N_{pe}^\beta$ can be used and the parameters, $b$ and $\beta$, can be determined by fitting the experimentally measured Moon shadow shift. 

\begin{figure}[ht]
\centering
\includegraphics[scale=0.150]{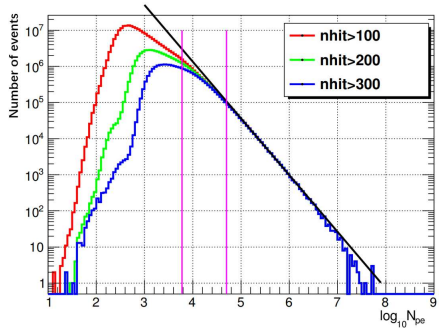}
\caption{The distribution of the total number of photo-electrons, $N_{pe}$, for shower events coming from a region around the Moon nominal position.
The histograms in red, green and blue correspond to events with number of hit cells,
$N_{hit}$, greater than 100, 200 and 300, respectively. The black solid line indicates a power law $\propto$ $N_{pe}^{-2.6}$, which fits the histograms for $N_{pe}> 50,000$. 
The two vertical magenta lines indicate the range used for the energy calibration using the Moon shadow.}
\label{Npe-distribution}       
\end{figure}

In Fig.~\ref{Npe-distribution} the distribution of $N_{pe}$ is shown for the events used in the Moon shadow analysis, for different values of the minimal number of detector units used in the reconstruction, $N_{hit}$, i.e., greater than 100, 200, or 300. Showers reconstructed in WCDA-1 using more than 200 units (hits) have a rather good directional resolution $0.39^{\circ}$~\cite{WCDA-on-Crab}. From Fig.~\ref{Npe-distribution} we can infer that above a certain energy, corresponding to $N_{pe}>50,000$, the showers are detected by WCDA-1 with rather high efficiency. As a matter of fact, the distribution of $N_{pe}$ has a clear power law behavior between 30,000 to 10$^6$ with a power index of -2.6, which is the same as the spectral index of the proton spectrum measured around 13 TeV by DAMPE~\cite{DAMPE-proton}.  This indicates that the efficiency does not change in this energy range, and $N_{pe}$ can be used as an energy estimator.

\subsection{Measurement of Moon shadow shifts for high energy showers }
The data set in this analysis was collected from 2019/05/01 to 2020/01/31. The total effective observation time to the Moon with zenith angles smaller than $45^{\circ}$ is 731.2 hours. To calculate the deflection angle as a function of the energy, events are grouped into five groups according to $N_{pe}$ as shown in Table 1. During the significance analysis, only events with zenith angles $<$ $45^{\circ}$ and $N_{hit}$ $>$ 200 are used. 

For the data set in each group, the sky map in celestial coordinates near the moon region is divided into a grid of $0.02^{\circ} \times 0.02^{\circ}$ bins and filled with detected events according to their reconstructed arrival directions. The number of events in each grid denoted as $n$. The number of background events in each grid is estimated by the equal zenith method~\cite{Amenomori}, denoted as $b$. The deficit significance in each grid is estimated by Li \& Ma formula~\cite{lima}.  The maximum deficit significance around the Moon region in each $N_{pe}$ group are also shown in Table 1, where the corresponding Moon shadow westward shifts, uncertainties and  are also listed. 

\begin{figure}[ht]
\centering
\includegraphics[scale=0.30]{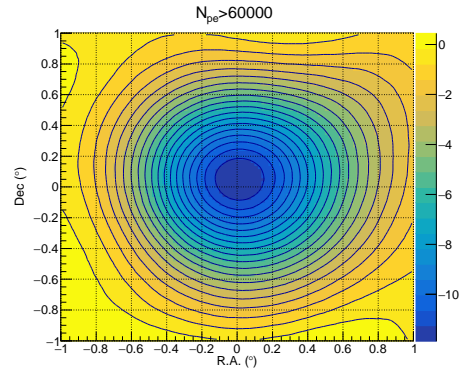}
\caption{The significance map of the Moon shadow for shower events detected by WCDA-1 with a $N_{pe}$ greater than 60,000. 
The deflection of the CRs by the GMF, at such high energies, is seen to be quite small.}
\label{Moon-shadow-overall}       
\end{figure}

For the $N_{pe}$ group with $N_{pe} > 60,000$, a significance as high as 10.9~$\sigma$ can be achieved. In Fig.~\ref{Moon-shadow-overall}, the significance of the deficit is plotted as a function of the arrival direction in right ascension (RA) and declination (Dec). The significance map can be fitted by a bi-dimensional Gaussian function to determine the location of the shadow center.
Its shift with respect to the nominal Moon position is $0.02^\circ\pm0.03^\circ$ in the Dec direction , while is quite small in the RA direction, i.e, $0.005\pm0.03$. 
The statistical uncertainty is the dominant contribution given the limited statistics. 
The standard deviation $\sigma$ of the Gaussian function is found to be $0.33^\circ\pm0.05^\circ$, which is the result of  the combination of the size of the Moon (angular extension 0.25$^\circ$) and the angular resolution of the detector~\cite{WCDA-on-Crab}.
The deficit number of events contained in the angular region $\sigma_{DEC}^2+\sigma_{RA}^2<2\sigma^2$ is about 63\% of the total deficit number, a value that is consistent with the simulation.
For lower energies, the shadow shifts towards the West and its size increases as well, indicating a worsening of the angular resolution.

\subsection{Calculation of Moon shadow shifts by raytracing in GMF}
\begin{figure}[h]
\centering
\includegraphics[scale=0.3]{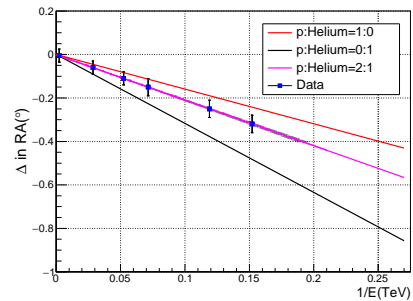}
\caption{The deflection angle of westward Moon shadow shift is graphed versus the inverse of the CR particle energy for pure protons (red line) and pure Helium nuclei (black line) according to the ray-tracing simulations. The blue squares, which are the measured mean shift of the Moon shadow with its statistical error in the corresponding $N_{pe}$ bins, lie along the  pink line representing the mixture of  protons and Helium nuclei with ratio of 2.0:1. The uncertainty in the composition is represented by the shaded area around the pink line, which is so small that it is barely visible.}
\label{shift-Npe}       
\end{figure}

\begin{figure}[ht]
\centering
\includegraphics[scale=0.3]{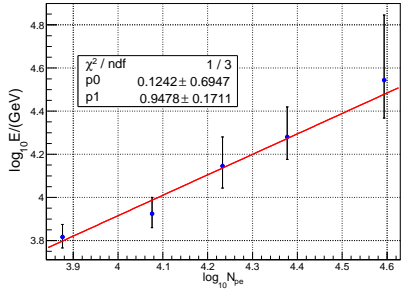}
\caption{The average shower energy measured using the Moon shadow shift versus $N_{pe}$, the total number of photo-electrons detected by WCDA-1 detector. }
\label{E-Npe}       
\end{figure}

The expected Moon shadow shift westward has been calculated by using a ray-tracing simulation which propagates protons and Helium nuclei coming from the Moon direction through GMF. The Interplanetary Magnetic Field, due to the solar wind, is by far less intense and can be neglected.  The Tsyganenko-IGRF model~\cite{GMF-model} has been used to describe the magnetic field in the Earth-Moon system. We find that the displacement obtained applying this model to the propagation of protons and Helium nuclei  can be represented  above 5 TeV  by the simple dipole approximation
\begin{equation}
    \Delta = 1.59^{\circ}/R(TV)
\end{equation}
where R(TV) is the particle rigidity  E(TeV)/Z. Thus the expected shift for Helium nuclei is a factor of 2 greater than the shift of protons of the same energy. However, for a given energy, the shower size $N_{e}$ of Helium nuclei is less than the size of proton-induced shower. According to the simulation, the energy of an Helium nuclei generating at the LHAASO altitude a shower size equal to the size of a proton of energy $E_{P}$ is $E_{He} \approx 1.9E_{P}$.  The rigidity of the Helium nuclei with the same shower size $N_{e}$ of proton is then $R \approx 1.9E_{P}/2$ , very close to the proton rigidity. This result has been already obtained in the Moon shadow analysis performed by ARGO-YBJ~\cite{ARGO-YBJ}. Thus, at a given selected interval of $N_{pe}$, corresponding to a shower size $N_{e}$ interval, the displacement of Helium-induced showers is practically equal to that of showers generated by proton primaries. Moreover, fixing a $N_{pe}$ interval we select a primary cosmic ray beam with a fraction Helium nuclei : protons is 2. after triggering
assuming the same flux for both components as measured by the CREAM experiment in this energy range~\cite{CREAM-proton-helium}. The mean energy of this beam is  $E_{u} \approx 1.3 E_{P}$. Since the westward shift of these particles is equal to the displacement of protons of energy $E_{P}$, that is $\Delta$ = 1.59/$E_{P}$ = k/$E_{u}$, we find the relation  $\Delta \approx 2.1/E_{u}$. This result is  fully confirmed by a detailed  simulation which provides the amount of the shift  $\Delta = ( 2.1 \pm 0.3)/E$  where E is the median energy of the cosmic beam selected by fixing $N_{pe}$. This relation is shown in Fig.~\ref{shift-Npe} where the deflection angles of proton and Helium nuclei  pure beams are also reported for comparison.  Using the results of Fig.6 we can attribute the cosmic ray median energy to each $N_{pe}$ selected interval reported in Table 1. Therefore, the energy scale can be fixed in the $N_{pe}$ range  6,000- 60,000 as shown in Fig.~\ref{E-Npe} . In the energy range from 6.6 TeV to 35 TeV $N_{pe}$ can be used as energy proxy according to the relation $E[GeV]=bN_{pe}^\beta$, $\beta=0.95\pm0.17$ and $b=1.33^{+5.26}_{-1.06}$.

.

\begin{table}
  \centering
  \caption{Moon shadow  shifts in RA, the significance of the Moon shadow, and the reconstructed median energy with its error, grouped bins of $N_{pe}$.}
  \label{Npe-bins}
  \small
  \begin{tabular}{lcccc}
    \hline
     Range of N$_{pe}$ & Shift of the &Significance & Median E  \\
     &  Moon shadow  ($^\circ$)&$(\sigma$) &(TeV)\\
    \hline
    6,000-10,000  &-0.32$\pm0.04$   &18.2 & 6.6$\pm0.8$  \\
    10,000-15,000  &-0.25$\pm0.04$  &14.0 & 8.4$\pm1.3$ \\
    15,000-20,000   &-0.15$\pm0.04$ &11.6& 14.0$\pm3.7$ \\
    20,000-30,000   &-0.11 $\pm0.03$  &11.9& 19.1$\pm5.2$ \\
    30,000-60,000   &-0.06 $\pm0.03$  &10.8& 35$\pm17.5$\\
    $>$60,000        &-0.005$\pm0.03$   &10.9& $>$50.0 \\
    \hline
  \end{tabular}
\end{table}

\subsection{Uncertainties of Energy Scale}
The uncertainty of the energy scale is due mainly to statistics, which are caused by the uncertainty of the measurements of Moon shadow shifts. The uncertainty of the Moon shadow shift can be transferred to the energy scale by using the fact that $d\Delta/\Delta=dE/E$. From Table \ref{Npe-bins} it can be seen that the average position of the shadow has an error of 0.04$^\circ$ in the lower energy bins, to become 0.03$^\circ$ at higher energies. 
These errors result in a rather large uncertainty in the energy scale reflected by the big error bars in Fig.~\ref{E-Npe}, i.e., from 12\% at 6.6 TeV to 50\% at 35 TeV. 
These uncertainties are expected to drop to 3\%  and  12\%, respectively, after four years of operations of the full WCDA detector.

Two systematic uncertainties may affect this analysis, one of them is the uncertainty of the ratio of the proton and Helium, whcich is found fluctating with an aplitude of 10\% over the energy range from 5 to 50 TeV according to the simulation. 
The other one is caused by the energy and angular resolution of the detector about 4\%. 

\section{Energy Scale of Shower Measurements for WFCTA}\label{sec:WFCTA}
In the previous section before the energy scale for WCDA-1 in the energy range from 6 TeV to 40 TeV has been determined. The correspondence between energy and its estimator $N_{pe}$ is modelled as  $E[GeV]$ = $1.33~N_{pe}^{0.95}$. 

The methods presented here can serve as a calibration of the energy estimation only in the  specific energy range used in the data analysis, while at higher energies, the energy scale has to rely solely on simulations of the air showers and detector response.
\begin{figure}[ht]
\centering
\includegraphics[scale=0.32]{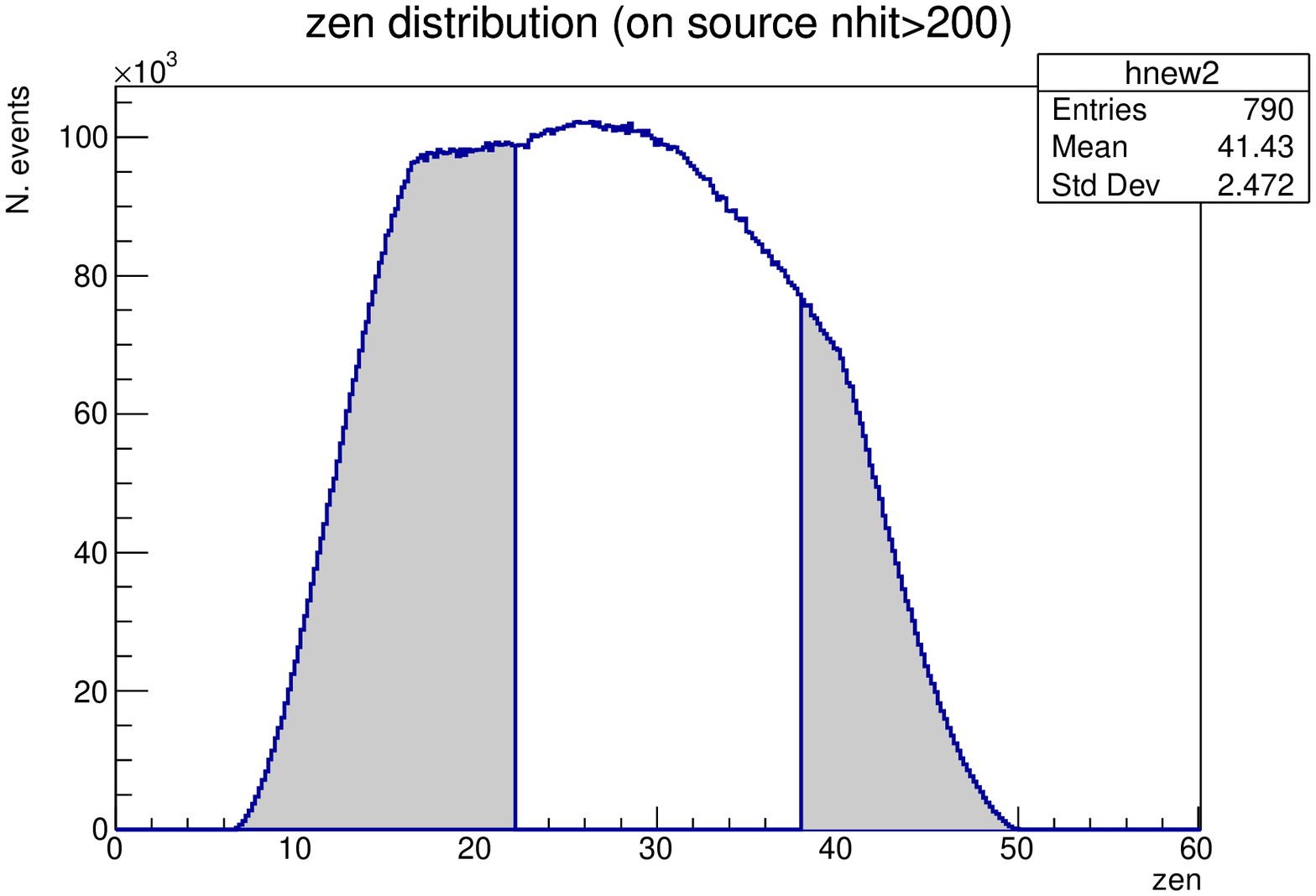}
\includegraphics[scale=0.32]{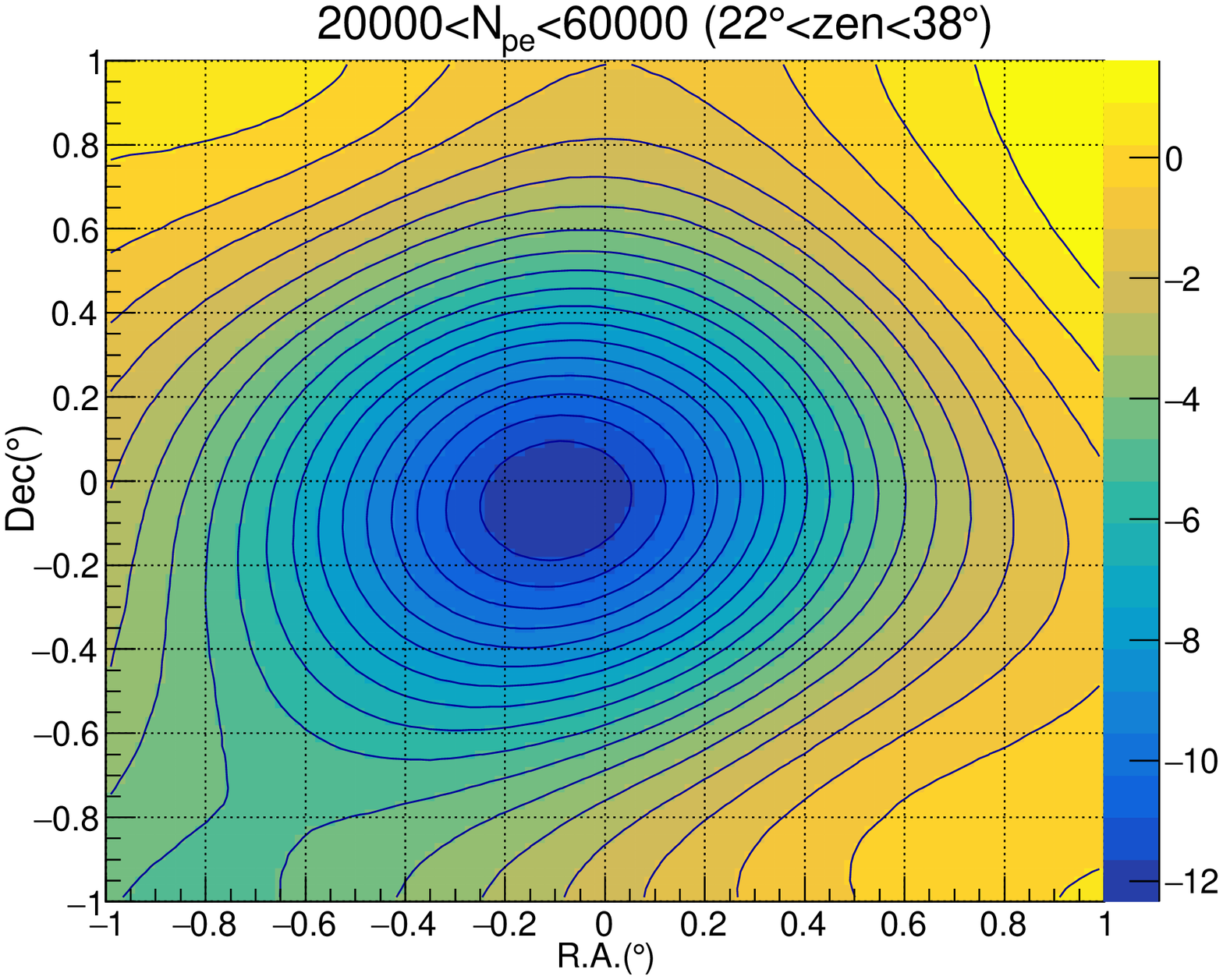}
\caption{(Upper) The distribution of zenith angles of all events used in the WCDA-1 Moon shadow analysis. The shaded areas represent the zenith range not covered by WFCTA telescopes. (Lower) The significance map of the Moon shadows, for events with $N_{pe}$ between 20,000 and 60,0000 and in the zenith angle range covered by the WFCTA telescopes, i.e., 22$^\circ\le\Theta_{Z}\le$ 38$^\circ$}
\label{zenith-distribution}       
\end{figure}

To detect the faint Cherenkov light, WFCTA telescopes have to be operated on dark nights or at most with partial moonlight, thus avoiding having the Moon in the FoV and making it impossible to measure Moon shadow shifts for determining the energy scale.
At the same time, the energy scale of WCDA-1 obtained with the Moon shadow shifts can be used to provide a reference for the energy calibration of WFCTA, using events triggered by both.  This cannot be done on an event by event basis, but instead with a set of commonly triggered events, which need therefore to have similar features with the set used for WCDA-1 analysis.

\begin{figure}[ht]
\centering
\includegraphics[scale=0.46]{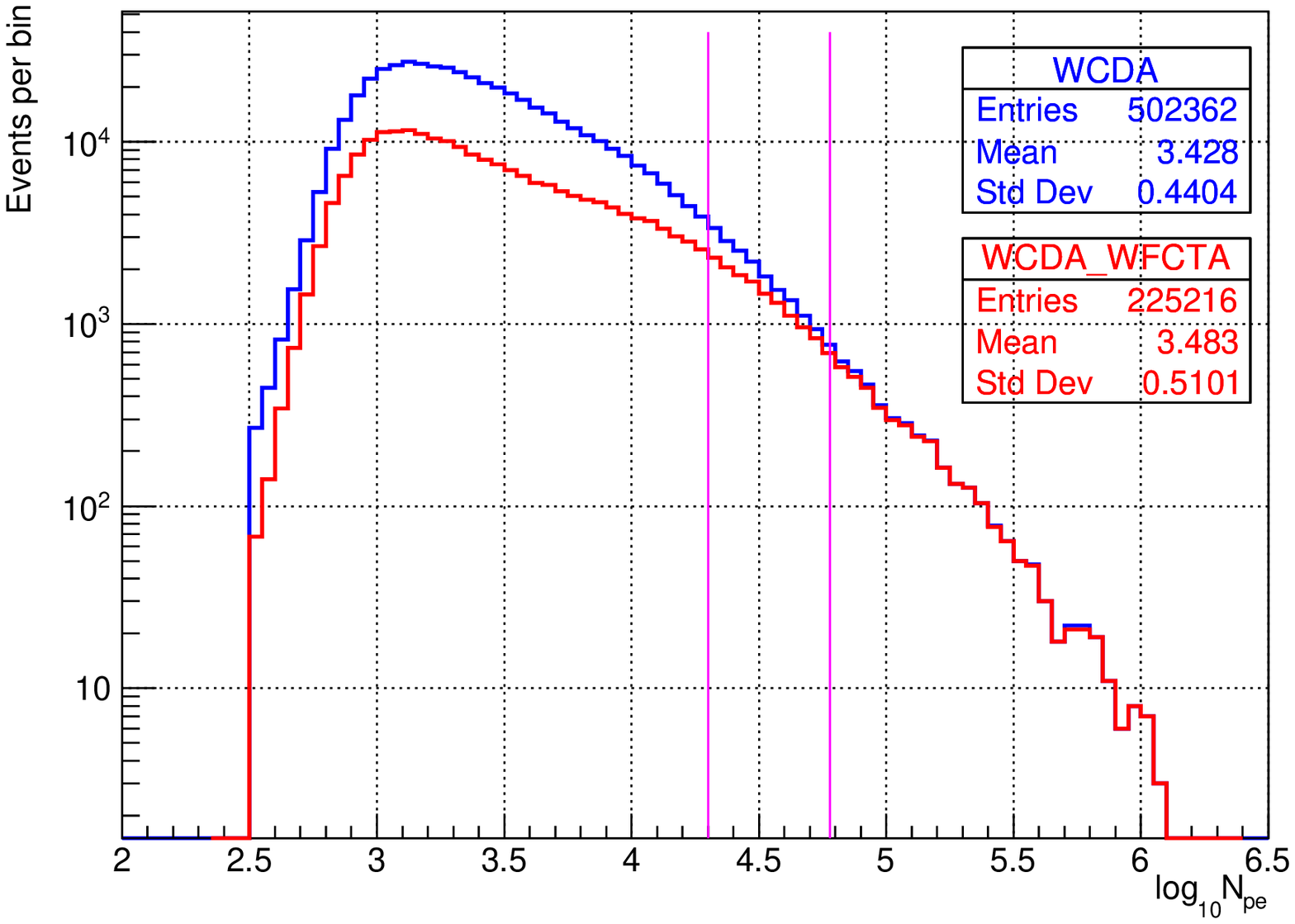}
\includegraphics[scale=0.35]{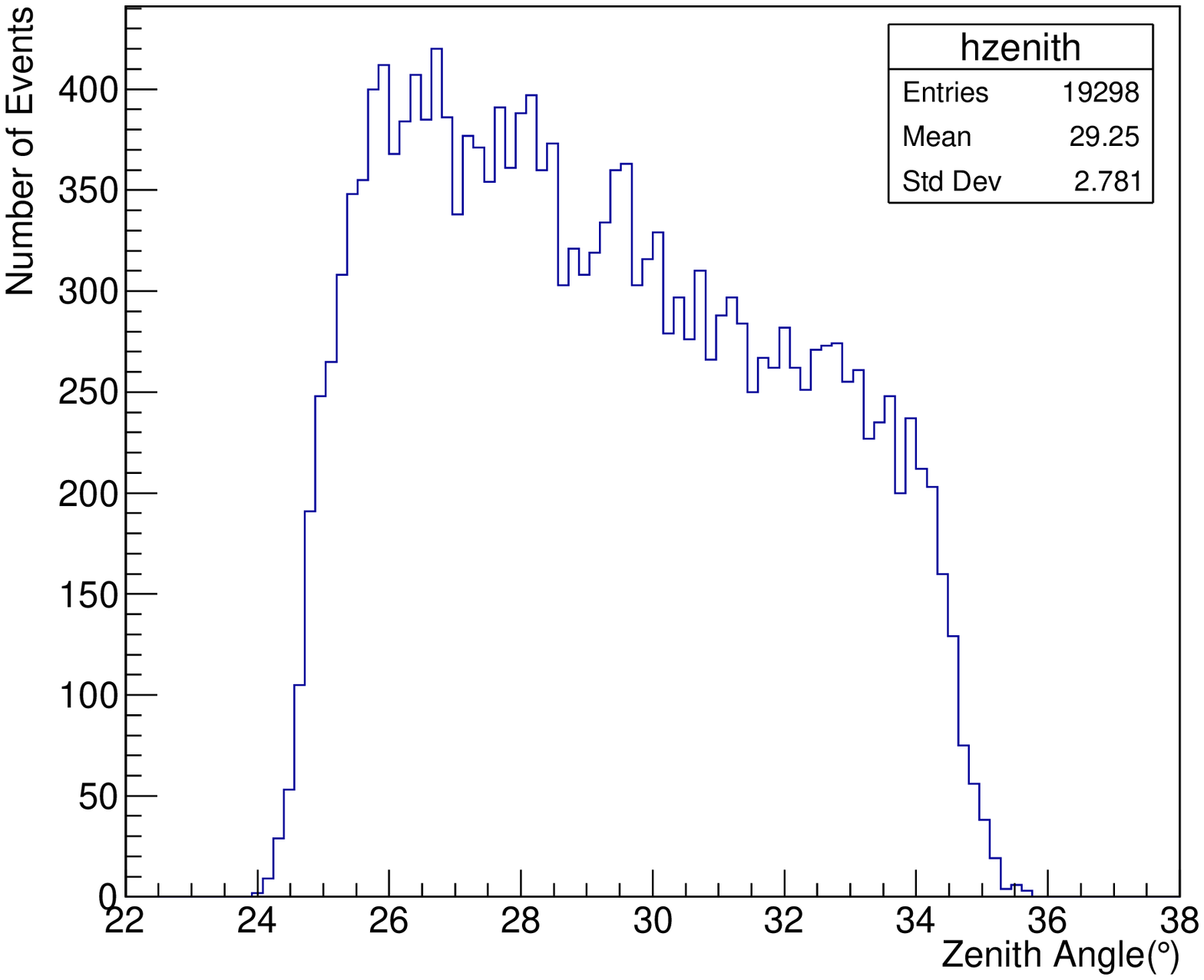}
\caption{The distributions of $N_{pe}$ (upper) and zenith angles (lower) for the events commonly triggered by both WCDA-1 and WFCTA. 
The distribution of $N_{pe}$ for WCDA-1 only (blue) is compared with that of common events used in the Moon shadow measurements described above (red). 
In the selected range of $N_{pe}$, indicated by the vertical lines, the selection efficiency is high enough to have almost the same ratio, proton : Helium nuclei = 2 : 1.}
\label{distributions-common-events}       
\end{figure}

The WFCTA telescopes are pointed at $30^{\circ}$ in zenith, which results in a zenith angle coverage from $22^{\circ}$ to $38^{\circ}$, taking into account the 16$^\circ$ FoV of the telescopes. While the events used to determine the WCDA-1 energy scale come from a region defined by the zenith angle range, i.e., mainly between $10^{\circ}$ to $45^{\circ}$ in section 4. Fig.~\ref{zenith-distribution} shows the zenith range covered by WCDA-1 and in white the zenith range covered by WFCTA.
Fig.~\ref{distributions-common-events} (upper) shows the distribution of  $N_{pe}$ for the events used in the Moon shadow analysis of WCDA (in blue) and 
the commonly triggered events (in red) obtained discarding the events falling in the shaded area as of Fig.~\ref{zenith-distribution}. Looking at the distribution in Fig.~\ref{distributions-common-events}, it can be seen that WCFTA triggers for 100\% of WCDA-1 events only for $N_{pe}$ above 60,000, and that the efficiency drops below 75\% for $N_{pe}$ below 20,000. This is due to the fact that the trigger efficiency of WCFTA telescopes decreases with increasing distance from the shower core and also for decreasing energy. On the contrary, the Moon shadow shift is better reconstructed for lower energies ($N_{pe}<$60,000) and when the core is well inside the ponds, i.e., further away from the telescopes.

Therefore, the common data set has been selected choosing the best compromise using the following criteria: $N_{hit}>200$, $20,000<N_{pe}<60,000$ and $22^{\circ}<\Theta_{z}<38^{\circ}$. 
Fig~\ref{distributions-common-events} (lower)
shows the distribution of zenith angles of commonly triggered events with the center of gravity if the images within 5$^{\circ}$ from the camera center.

Using these selection  criteria leads to a new significance map, shown in Fig.~\ref{zenith-distribution}, with the same analysis method as used for WCDA. 
The value obtained for the shift in this case is  $\Delta=(0.1\pm0.03)^{\circ}$, which translates to an energy 21$\pm 6.5$ TeV, using the formula $\Delta = -2.1/E$[TeV] as before. 
The uncertainty quoted is purely statistical and results in a relative error of the calibration of about 30\%. 
Clearly this result can be significantly improved over more years at collecting statistics. 

The density of Cherenkov photons produced during the shower development is proportional to the primary particle energy.
Therefore a good estimator of the energy  for WFCTA data is the sum of the number of photo electrons in the camera image, $\sum$, once corrected for geometric effects.
The quantity $\sum$ is usually called the \textit{size of the image}, in the traditional shower energy reconstruction scheme~\cite{ARGO-YBJ} for Cherenkov telescopes.
The correction on $\sum$ includes its dependence on the viewing angle $\beta$, the space angle between the shower direction and the main axis of the telescope, and the  perpendicular distance, $R_p$,  from the shower axis to the telescope in shower-detector-plane (SDP). 
The $R_P$ correction takes into account the fact that photon density decreases with increasing distance from the shower core.
The viewing angle $\beta$ correction is due to the weakening of the shower image on the camera for off-axis showers.
This is a combination of the shadow of the container in which the telescope is installed and also the reduced effective area of the mirror for off-axis showers.
The correction has been calculated with a detailed simulation of the WFCTA telescopes response~\cite{YLQ-composition-CPC} omitted here for the sake of brevity. 

The shower energy can be estimated with corrected $\sum$  with a resolution of 20\%, for energies below 100 TeV, and 15\% above 100 TeV with a systematic shift less than 5\%~\cite{YLQ-composition-CPC}.

\begin{figure}[ht]
\centering
\includegraphics[scale=0.35]{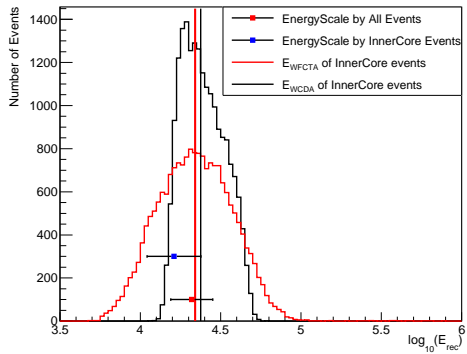}
\caption{The distributions of reconstructed energies. The black line indicates the distribution of energies reconstructed by WCDA-1 for the commonly triggered events, using the estimator $E[GeV]=1.33\cdot N_{pe}^{0.95}$. The vertical black line indicates the median energy (23.6 TeV) of the distribution. The red line indicates energies reconstructed by WFCTA telescopes assuming a primary composition ratio of protons : Helium nuclei = 1 : 1. The vertical red line indicates the mean energy (21.9 TeV) of the distribution. The energy scale with it uncertainty (21$\pm$6.5 TeV) measured using the Moon shadow shift with available data (red square). The blue square indicates the energy scale estimated only for events with a core located inside WCDA-1.}
\label{WFCTA-Energy} 
\end{figure}

In order to correctly reconstruct the energy, a proper set of data has been selected.
For WCDA events, apart from the standard cut $20,000<N_{pe}<60,000$, a requirement is added to have the shower core to be well inside the pond, i.e., $|x_{core}|<$ 55 m and $|y_{core}|<$ 55 m . 
In addition, the WFCTA events are required to be not too far, i.e., \mbox{$R_{p}<$ 100 m},  to have at least 2 triggered pixels, and to have the center of gravity of the images  within $5^{\circ}$ from the camera center.
The distribution of energies, $E_{\tiny WFCTA}$ reconstructed by $\sum$ for the sample selected with these cuts is shown in Fig.~\ref{WFCTA-Energy}. The mean value ( 21.9 TeV )of the distribution is shown by the vertical red line. The WCDA-derived energies of commonly triggered events with the same cuts can be reconstructed by the formula  $E[GeV]$ = $1.33~N_{pe}^{0.95}$, which is shown in Fig.~\ref{WFCTA-Energy} by the black line. The median energy (23.6 TeV) of the distributions is shown by the black vertical line.  

For comparison, the energy scale with its statistical uncertainty ( 21$\pm$6.5 TeV) determined from the Moon shadow shift are also reported in Fig.~\ref{WFCTA-Energy}. 
The agreement between the mean of the energy reconstructed by the WFCTA telescopes and the energy scale of WCDA-1 is evident and differences are of the order of 4\%, well below the 30\% uncertainty of the Moon shadow energy determined which is  dominated by the statistical error.

In order to evaluate the bias introduced by the requirement to have the shower core well within WCDA, the energy scale has been estimated by the Moon shift method adding the conditions $|x_{core}|<$ 55 m and $|y_{core}|<$ 55 m to the previous ones,i.e., $N_{hit}>200$, $20,000<N_{pe}<60,000$ and $22^{\circ}<\Theta_{z}<38^{\circ}$.
The resulting deflection angle is  $\Delta=(0.13\pm0.05)^{\circ}$, which translates into an energy of 16.2$\pm$6.2 TeV, according to the formula $\Delta = -2.1/E$(TeV), this is shown as a blue square with its error, corresponding to a relative error of 38\% in Fig.~\ref{WFCTA-Energy}.
The change of the energy scale introduced by the further require of having the shower core well within WCDA is about 30\%.

\begin{figure}[ht]
\centering
\includegraphics[scale=0.3]{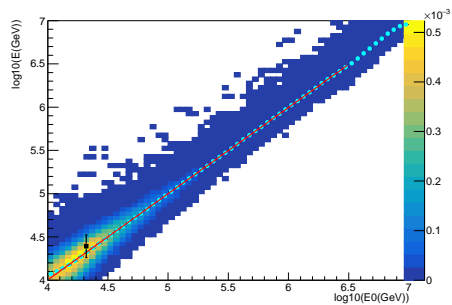}
\caption{Distributions of $E$ reconstructed by WFCTA and primary energy $E_0$ for primary composition ratio  protons : Helium nuclei = 1 : 1 in the energy range form 10 TeV to 10 PeV.  The black square indicates the energy scale (21 TeV) obtained from the Moon shadow shift and the mean energy as reconstructed by WFCTA telescopes. The Color scale indicates the relative number of events in each bin. }
\label{E_televsE_0}       
\end{figure}

To extrapolate the WFCTA energy scale as determined in the overlapping region with WCDA,  the correspondence between the energy estimator of WFCTA and the primary energy  has been checked using simulations.
Fig.~\ref{E_televsE_0} shows the distribution of  $E_{\tiny WFCTA}$ reconstructed by WFCTA as a function of input primary energy $E_0$, spanning a range from 30 TeV to 10 PeV.
In addition, the energy scale determined from Moon shadow shifts in WCDA is graphed as a black square together with the error bar at $21.0\pm6.5$ TeV. 
The uncertainty here is dominated by the statistical errors, as the systematic error coming from composition uncertainty, about 3\%, contribute marginally.

\section{Conclusion}\label{sec:conclusions}
In this work, we have shown how the shift of Moon shadow in cosmic rays,  due to the geomagnetic field, can be used to establish the energy scale of the WCDA detector in the range  from 6 TeV to 35 TeV, using the estimator $E[GeV]=1.33\cdot N_{pe}^{0.95}$, based on the total number of photo-electrons $N_{pe}$ measured in the pond.
The uncertainty of the energy scale varies from $\pm12\%$ at 6.6 TeV to $\pm50\%$ at 35 TeV and is dominated by statistical errors. The systematic error coming from the assumption of the composition of the primary ratio protons : Helium nuclei with respect with the direct measurement by space born experiments~\cite{CREAM-proton-helium} has been shown to be about 3\%.


Given the impossibility to measure the Moon shadow directly with WFCTA, a set of events commonly triggered with WCDA-1 have been used to correlate the energy scale of  WCDA to the measurement using WFCTA telescopes.  
Unfortunately, the two types of detector have different energy thresholds, and this overlap occurs at upper limited energy range of WCDA, where the Moon shadow shift method is less reliable.
This results in a rather large uncertainty of about $30\%$ on the 'common' energy scale. As a matter of fact, this uncertainty is largely dominated by the low statistics of event, given the small contribution of the systematic error from the lack of information on the composition in this energy. Other systematic uncertainties related to the hadronic interaction model used in the simulation, at this energies, should be smaller than the previous one. 

In few years LHAASO will accumulate huge statistics, which will allow us to reduce the uncertainty of the energy scale below 12\%.
For  showers with energy  above 35 TeV, the energy reconstruction  totally relies on the simulation of the WCFTA, which indicates that the energy estimator $\sum$ will provide a good energy resolution.

\vspace*{1cm}

\section{Acknowledgments}
The authors would like to thank all staff members who work at the LHAASO site all year around to keep the system running efficiently and smoothly, even in the demanding conditions at a mean altitude of 4400 meters above sea level. We are grateful to the Chengdu Management Committee of Tianfu New Area for the constant financial support to the research with LHAASO data. This research work is also supported by the National Key R\&D program of China, with the grant 2018YFA0404201, 2018YFA0404202 and  2018YFA0404203, the National Natural Science Foundation of China, with NSFC grants 11635011, 11761141001, 11905240, 11503021, 11205126, 11947404, 11675187, U1831208, Schools of Science and Technology Plan from SiChuan Province  grant No.20SYSX0294, and Thailand Science Research and Innnovation grant RTA6280002.

\end{document}